\documentclass[runningheads]{llncs}
\usepackage[T1]{fontenc}
\pdfoutput=1
\usepackage{xcolor, soul}
\usepackage[misc]{ifsym}
\usepackage{graphicx}
\usepackage{booktabs}
\usepackage{siunitx}
\title{
  A New Dataset and A Baseline Model for
  Breast Lesion Detection in Ultrasound Videos
}

\titlerunning{
  A New Dataset and CVA-Net for
  Breast Ultrasound Videos Detection
}

\authorrunning{Z. Lin et al.}
\author{
  Zhi Lin\inst{1}$\ ^{\star}$ \and
  Junhao Lin\inst{1}
  \thanks{Z. Lin and J. Lin---Equal contribution.} \and
  Lei Zhu\inst{2,3}$^{\textrm{(\Letter)}}$ \and
  Huazhu Fu\inst{4} \and \\
  Jing Qin\inst{5} \and
  Liansheng Wang\inst{1}
}
\institute{
  Department of Computer Science, School of Informatics, \\
    Xiamen University, Xiamen, China \and
  ROAS Thrust, System Hub, 
    The Hong Kong University of Science and \\
    Technology (Guangzhou), Guangzhou, China \\
    \email{leizhu@ust.hk} \and
  Department of Electronic and Computer Engineering, The Hong Kong University of \\
    Science and Technology, Hong Kong SAR, China \and
  Institute of High Performance Computing, Agency for Science, \\
    Technology and Research, Singapore \and
  School of Nursing, The Hong Kong Polytechnic University, 
    Kowloon, Hong Kong
}

\begin{document}

\maketitle
\raggedbottom

\begin{abstract}
Breast lesion detection in ultrasound is critical for breast cancer diagnosis.
Existing methods mainly rely on individual 2D ultrasound images or combine unlabeled video and labeled 2D images to train models for breast lesion detection. 
In this paper, we first collect and annotate an ultrasound video dataset (188 videos) for breast lesion detection.
Moreover, we propose a clip-level and video-level feature aggregated network (CVA-Net) for addressing breast lesion detection in ultrasound videos by aggregating video-level lesion classification features and clip-level temporal features.
The clip-level temporal features encode local temporal information of ordered video frames and global temporal information of shuffled video frames.
In our CVA-Net, an inter-video fusion module is devised to fuse local features from original video frames and global features from shuffled video frames, and an intra-video fusion module is devised to learn the temporal information among adjacent video frames.
Moreover, we learn video-level features to classify the breast lesions of the original video as benign or malignant lesions to further enhance the final breast lesion detection performance in ultrasound videos.
Experimental results on our annotated dataset demonstrate that our CVA-Net clearly outperforms state-of-the-art methods.
%%
%% We shall release our annotated dataset, our code, and our results upon the publication of this work.
The corresponding code and dataset are publicly available at \url{https://github.com/jhl-Det/CVA-Net}.
\end{abstract}

\section{Introduction}
\label{sec:introduction}

Breast cancer is a leading cause of death for women worldwide~\cite{wild2020world}.
Currently, ultrasound imaging is the most commonly used and effective technique for breast cancer detection due to its versatility, safety, and high sensitivity~\cite{stavros1995solid}.
Detecting breast lesions in ultrasound is often taken as an important step of computer-aided diagnosis systems to assist radiologists in the ultrasound-based breast cancer diagnosis~\cite{yap2017automated,zhu2020second,chen2019semi}.
However, accurate breast lesion detection in ultrasound videos is challenging due to blurry breast lesion boundaries, inhomogeneous distributions, changeable breast lesion sizes and positions in dynamic videos.

Existing methods~\cite{yang2020temporal,movahedi2020automated,zhang2020birads,qi2019automated,xue2021global}  mainly performed the breast lesion segmentation or detection in 2D ultrasound images, or fused unlabeled videos with labeled 2D images for ultrasound video breast lesion detection. 
With the dominated results of convolutional neural networks on medical imaging, it is highly desirable to extend deep-learning-based breast lesion detection from the image level to video level, since the latter can leverage temporal consistency to address many in-frame ambiguities.
The major obstacle for this extension is the lack of an ultrasound video dataset with appropriate annotations for breast lesion segmentation, both of which are essential for training deep models for breast lesion segmentation in ultrasound videos. 

\begin{figure*}[!t]
\centering
\includegraphics[width=0.8\textwidth]{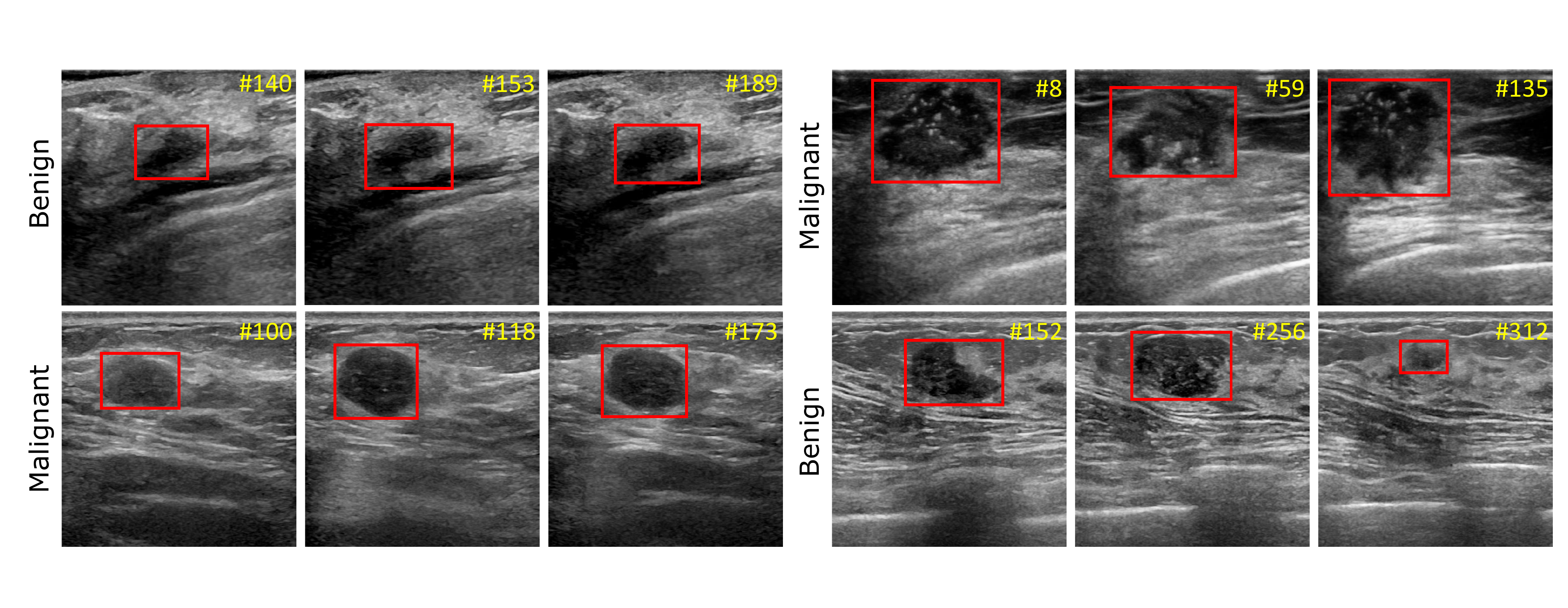}
\vspace{-3mm}
\caption{
  Examples of our annotated ultrasound video dataset for breast lesion detection.
}
\vspace{-5mm}
\label{fig:dataset}
\end{figure*}

To do so, we in this work first provide a video dataset for breast lesion detection in ultrasound; see Fig.~\ref{fig:dataset} for several examples of annotated videos.
Then, we present a novel network for boosting breast lesion detection in ultrasound videos by aggregating video-level classification features and clip-level temporal features, which contains a local temporal feature from the input video frames and a global temporal information from shuffled video frames. 
The contributions of this work could be summarised as:
1) We develop a novel network to learn clip-level temporal features and video-level lesion classification features for boosting breast lesion detection in ultrasound videos. 
2) We collect and annotate a video dataset (118 videos) for breast lesion detection in ultrasound videos. 
3) An inter-video fusion module is devised to attentively aggregate local features from the original video frames and global features from the shuffled video frames, while an intra-video fusion module is developed to fuse temporal features encoded among adjacent video frames.
4) Experimental results on our annotated dataset demonstrate that our network sets a new state-of-the-art performance on breast lesion detection in ultrasound videos.

\section{Method}
\label{sec:method}

\begin{figure*}[!t]
\centering
\includegraphics[width=\textwidth]{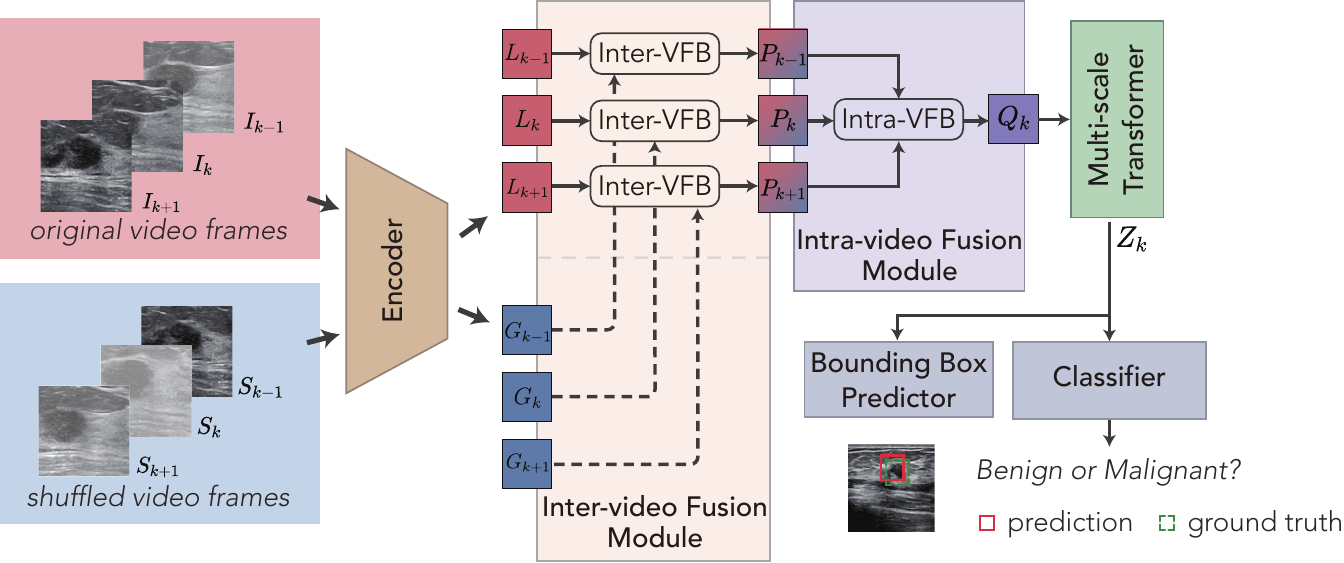}
\vspace{-3mm}
\caption{
  Schematic illustration of our clip/video-level feature aggregation network (CVA-Net) for breast lesion detection network in ultrasound videos.
  ``Inter-VFB'' and ``Intra-VFB'' denotes a inter-video fusion block and a intra-video fusion block.
}
\label{fig:pipeline}
\end{figure*}

Fig.~\ref{fig:pipeline} shows the schematic illustration of the developed clip-level and video-level feature aggregation network (CVA-Net).
The motivation behind our CVA-Net is to integrate video-level lesion classification features and clip-level features from adjacent video frames, and such clip-level features include local temporal information from the original video and global temporal information from the shuffled video.
To do so, given an input ultrasound video with $T$ frames, we first shuffle the ordered frames index sequence $\{1,\cdots,T\}$ of the input video to obtain a new index sequence, which is then used to generate a shuffled video.
For a current video frame ($I_k$), our CVA-Net takes three neighboring images (denoted as $I_k$, $I_{k-1}$, and $I_{k+1}$) of the input video, and then passes $I_k$, $I_{k-1}$, and $I_{k+1}$ into a feature extraction backbone (i.e., ResNet50~\cite{2016Deep}) with three convolutional layers to obtain three features, which are denoted as $L_k$, $L_{k-1}$, and $L_{k+1}$).
Apparently, $L_k$, $L_{k-1}$, and $L_{k+1}$ contain three CNN feature maps with different spatial resolutions.
Meanwhile, we take three corresponding images (denoted as  $S_k$, $S_{k-1}$, and $S_{k+1}$) of the shuffled video, apply data augmentation techniques (e.g., horizontal flip, random crop, random pepper) on them, and pass augmented images into the feature extraction backbone to produce another three features $G_k$, $G_{k-1}$, and $G_{k+1}$.
Afterwards, we devise an inter-video fusion module to integrate local temporal features ($L_k$, $L_{k-1}$, and $L_{k+1}$) from input ultrasound video, and global features ($G_k$, $G_{k-1}$, and $G_{k+1}$) from the shuffled video to obtain three features, which are denoted as $P_k$, $P_{k-1}$, and $P_{k+1}$.
Then, we devise an intra-video fusion to fuse these three features $P_k$, $P_{k-1}$, and $P_{k+1}$ to produce a new feature map $Q_k$, which is then passed into a classifier to predict whether the breast lesions are benign or malignant, and a bounding box predictor to produce the final breast lesion detection of the current video frame $I_k$.

\subsection{Inter-video Fusion Module}

Existing breast lesion detection in ultrasound video methods~\cite{chen2019semi} extracted the temporal information encoded in ordered video frames sampled from the input video~\cite{chen2020memory}.
Unlike this, we devise an inter-video fusion module to utilize three frames from a shuffled video to remove the temporal order and enhance the global semantic information for detecting breast lesions.
As shown in Fig.~\ref{fig:pipeline}, our inter-video fusion module has three inter-video fusion blocks to fuse three pairs of local features and global features, which are $L_{k}$ and $G_{k}$, $L_{k-1}$ and $G_{k-1}$, as well as $L_{k+1}$ and $G_{k+1}$, respectively.
By doing so, we can fuse the local information from the input video and the global information from the shuffled video, thereby improving the video breast lesion detection.

\begin{figure*}[!t]
\centering
\includegraphics[width=\textwidth]{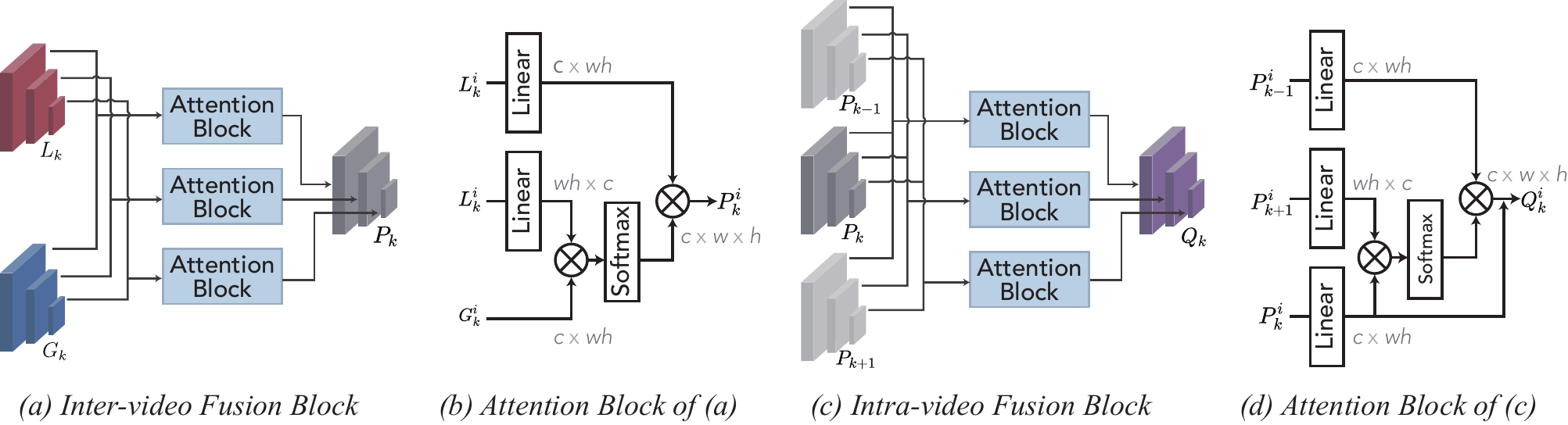}
\vspace{-5mm}
\caption{
  Schematic illustration of the inter-video fusion module and intra-video fusion module of our network (see Figure~\ref{fig:pipeline}).
}
\vspace{-3mm}
\label{fig:module}
\end{figure*}

Figure~\ref{fig:module}(a) depicts the schematic illustration of the inter-video fusion block, which integrates the local feature $L_{k}$=($L_{k}^1$, $L_{k}^2$, $L_{k}^3$) and the global feature $G_{k}$=($G_{k}^1$, $G_{k}^2$, $G_{k}^3$).
The inter-video fusion block utilizes three attention blocks to integrate three features of $L_{k}$ and $G_{k}$, and the $i$-th attention is for $L_{k}^i$ and $G_{k}^i$ (i=1,2,3); see Figure~\ref{fig:module}(b).
Specifically, we first apply two different linear layers on $L_{k}^i$, reshape both of them into two $c\times wh$ matrices, and also reshape $L_{k}^i$ into a $wh\times c$ matrix. Then we multiply two reshaped matrices to obtain a similarity matrix ($wh\times wh$), followed by a softmax layer.
Then, we multiple the result with another reshaped matrix of $L_{k}^i$, and reshape the multiplication output to obtain the output feature $H_{k}^i$ ($c\times w\times h$). 
Mathematically, $P_{k}^i$ is computed by:
\begin{equation}
\label{inter_attention}
    P_{k}^i = \hat{L}_k^i \times \textit{Softmax} \left( \mathcal{R}({G}_k^i) \times \mathcal{R}(\tilde{L}_k^i) \right)   \ ,
\end{equation}
where $\mathcal{R}$ denotes a reshape operation. 
$\tilde{L}_k^i$ and $\tilde{L}_k^i$ represents the obtained features via utilizing two different linear layers on the input feature ${L}_k^i$.

\subsection{Intra-video Fusion Module}

Although our inter-video fusion module is capable of integrating several paired of local features from the original video and the global features of the shuffled video, it neglects the temporal information among the adjacent video frames.  
In this regard, we develop an intra-video fusion module to capture the temporal feature among adjacent video frames. 
Figure~\ref{fig:module}(c) shows the details of our intra-video fusion module, which further integrates three output features ($P_{k-1}$, $P_{k}$, and $P_{k+1}$) of inter-video fusion from three adjacent video frames. 
Let $P_{k}$=($P_{k}^1$, $P_{k}^2$, $P_{k}^3$),
$P_{k-1}$=($P_{k-1}^1$, $P_{k-1}^2$, $P_{k-1}^3$), and $P_{k+1}$=($P_{k+1}^1$, $P_{k+1}^2$, $P_{k+1}^3$) to denote three CNN features of $P_{k-1}$, $P_{k}$, and $P_{k+1}$.
To do so, our intra-video fusion module utilizes three attention blocks to integrate different CNN features, and $i$-th attention block fuses $P_{k-1}^i$, $P_{k}^i$, and $P_{k+1}^i$.
Specifically, Figure~\ref{fig:module}(d) shows the schematic illustration of the $i$-th attention block.
We first apply two linear layer on two input features $P_{k}^i$ ($c\times w\times h$), and $P_{k+1}^i$ ($c\times w\times h$), and then reshape both of them to two matrices, and their sizes are ($c\times wh$) and ($wh \times c$).
We then multiply these two reshaped matrices to obtain a $wh \times wh$ similarity matrix, which is passed into a Softmax layer to fuse two input features $P_{k}^i$ and $P_{k+1}^i$ .
Meanwhile, we utilize a linear layer on another input feature $P_{k-1}^i$ ($c\times w\times h$) and reshape it into $c \times wh$.
To further fuse $P_{k-1}^i$, we multiply its reshaped matrix with the similarity matrix from $P_{k}^i$ and $P_{k+1}^i$ to obtain a new feature ($c \times wh$), which is then reshaped into a 3D feature map ($c \times w \times h$); see $Q_{k}^i$ of Figure~\ref{fig:module}(d).
Mathematically, $Q_{k}^i$ is computed by:
\begin{equation}
\label{intra_attention}
    Q_{k}^i =  \mathcal{R}(\hat{P}_{k-1}^i) \times \textit{Softmax}\left( \mathcal{R}(\hat{P}_k^i) \times \mathcal{R}(\hat{P}_{k+1}^i) \right) \ ,
\end{equation}
where $\mathcal{R}$ denotes a reshape operation. 
$\hat{P}_k^i$, $\hat{P}_{k+1}^i$, and $\hat{P}_{k-1}^i$ represent the obtained features via utilizing a linear layer on three input features $P_k^i$, $P_{k-1}^i$, and $P_{k+1}^i$.

\subsection{Our Network}

Apparent from learning clip-level features (i.e., $P_{k}$=($P_{k}^1$, $P_{k}^2$, $P_{k}^3$)) via fusing local and global temporal features via our inter-video fusion modules and our intra-video fusion module, we further learn a video-level information, which represents the classification result (benign or maligant) on each breast lesion of the input ultrasound video.
To do so, we first pass three features (i.e., $P_{k}^1$, $P_{k}^2$, and $P_{k}^3$) of $P_{k}$ into a multi-scale transformer (i.e., deformable-DETR~\cite{zhu2020deformable}) to learn a long-range dependency feature map $Z$.
We then apply a linear layer on $Z$ to classify the breast lesions as benign lesions or malignant lesions.
Furthermore, we predict bounding boxes of breast lesion as the final breast lesion detection result of the current video frame $I_k$ (see Fig.~\ref{fig:pipeline}).
We utilize a negative log-likelihood to compute the lesion classification loss, and a linear combination of the $ {\ell}_1 $ loss and the generalized IoU loss~\cite{Rezatofighi2019GeneralizedIO} for computing the breast lesion detection loss.

\vspace{.5mm}
\noindent
\textbf{Implementation Details.} \
We utilize ResNet-50~\cite{2016Deep} pre-trained on ImageNet~\cite{2009ImageNet} to initialize the backbone of our network, and deformable~DETR~\cite{zhu2020deformable} pre-trained on COCO~2017~\cite{lin2014microsoft} to initialize the multi-scale transformer module of our network, while
other network parameters are initialized by a normal distribution.
All training videos are randomly horizontally flipped, resized, and cropped for data augmentation.
%%~\cite{kingma2014adam}
Our network is implemented on Pytorch and trained using an Adam with 50 epochs, an initial learning rate of \num{2e-4}, and a weight decay of \num{1e-4}.
The whole architecture is trained on three GeForce RTX 2080 Ti GPUs, and each GPU has a batch-size of 1.

\section{Experiments and Results}
\label{sec:experiments}

\noindent
\textbf{Dataset.} \ 
Note that there is no public ultrasound video breast lesion detection benchmark dataset with annotations. 
% To evaluate the effectiveness of the developed network, we collect a breast lesion ultrasound video dataset with $188$ videos, which has $113$ malignant videos and $75$ benign videos.
%\hl
{To evaluate the effectiveness of the developed network, we collect a breast lesion ultrasound video dataset with $188$ videos, which has $113$ malignant videos and $75$ benign videos. These $118$ videos has $25,272$ images in total, and the number of ultrasound images at each video varied from $28$ to $413$.}
Each video has a complete scan of the tumor, from its appearance to the largest section, then to its disappearance.
All videos are acquired by LOGIQ-E9 and PHILIPS TIS L9-3.
Two pathologists with eight years of experience in breast pathology were invited to manually annotate the breast lesion rectangles inside of each video frame and give the corresponding classification label for the breast lesions of the video. 
We further randomly and evenly select $38$ videos from the dataset as a testing set (about 20\% of the dataset) and the remaining videos as the training set.

\vspace{.5mm}
\noindent
\textbf{Evaluation Metrics.} \ 
We utilize three widely-used metrics for quantitatively comparing different ultrasound video breast lesion detection methods.
They are average precision (AP), AP$_{50}$, and AP$_{75}$.
Please refer to~\cite{wang2020solo} for the definitions of AP, AP$_{50}$, and AP$_{75}$.

\begin{table}[!t]
\caption{
 Quantitative comparisons of our network and compared methods on our annotated video dataset.
}
\label{tab:sota}
\centering
\vspace{-3mm}
\begin{tabular}{l|c|c|ccc}
\bottomrule
  Method & Type & Backbone & AP & AP$_{50}$ & AP$_{75}$ \\
\hline
  GFL~\cite{li2020generalized} & image & ResNet-50 & 23.4 & 46.3 & 22.2 \\
  Cascade RPN~\cite{Vu2019CascadeRD} & image & ResNet-50 & 24.8 & 42.4 & 27.3\\
  Faster R-CNN~\cite{ren2015faster} & image & ResNet-50 & 25.2 & 49.2 & 22.3 \\
  VFNet~\cite{Zhang2021VarifocalNetAI} & image & ResNet-50 & 28.0 & 47.1 & 31.0 \\
  RetinaNet~\cite{lin2017focal} & image & ResNet-50 & 29.5 & 50.4 & 32.4 \\
\hline
  DFF~\cite{zhu2017deep} & video & ResNet-50 & 25.8 & 48.5 & 25.1 \\
  FGFA~\cite{2017} & video & ResNet-50 & 26.1 & 49.7 & 27.0 \\
  SELSA~\cite{Wu2019SequenceLS} & video & ResNet-50 & 26.4 & 45.6 & 29.6 \\
  Temporal ROI Align~\cite{gong2021temporal} & video & ResNet-50 & 29.0 & 49.9 & 33.1 \\
  MEGA~\cite{chen2020memory} & video & ResNet-50 & 32.3 & 57.2 & 35.7 \\
\hline
  CVA-Net (ours) & video & ResNet-50 & \textbf{36.1} & \textbf{65.1} & \textbf{38.5} \\
\toprule
\end{tabular}
\end{table}

\subsection{Comparisons with State-of-the-arts}

We compare our network against ten state-of-the-art methods, including five image-based methods and five video-based methods.
The image-based methods include, 
GFL~\cite{li2020generalized}, 
Cascade RPN~\cite{Vu2019CascadeRD},
Faster R-CNN~\cite{ren2015faster}, 
VFNet~\cite{Zhang2021VarifocalNetAI}, and 
RetinaNet~\cite{lin2017focal}.
The five video-based methods are 
DFF~\cite{zhu2017deep}, 
FGFA~\cite{2017}, 
SELSA~\cite{Wu2019SequenceLS}, 
Temporal ROI Align~\cite{gong2021temporal}, 
and MEGA~\cite{chen2020memory}. 
For providing a fair comparison, we obtain the detection results of all compared methods by exploiting its public implementations or implementing them by ourselves.
We also re-train these networks on our dataset, and ﬁne-tune the network parameters for achieving the best detection results of ultrasound video breast lesions. 

\vspace{.5mm}
\noindent
\textbf{Quantitative Comparisons.} \ 
Table~\ref{tab:sota} summarizes the quantitative results of our network and all ten compared breast lesion video detection methods.
In general, video-based methods tend to have larger AP, AP$_{50}$, and AP$_{75}$ scores than image-based ones.
Specifically, among ten compared methods, MEGA has the largest AP score of 32.3, the largest AP$_{50}$ score of 57.2, and the largest AP$_{75}$ score of 35.7. 
On the contrary, our method further outperforms MEGA in terms of all three metrics (AP, AP$_{50}$, and AP$_{75}$).
Compared to MEGA, our CVA-Net improves the AP score from 32.3 to 36.1, the AP$_{50}$ score from 57.2 to 65.1, and the AP$_{75}$ score from 35.7 to 38.5.

\begin{figure*}[!t]
\centering
\includegraphics[width=0.8\textwidth]{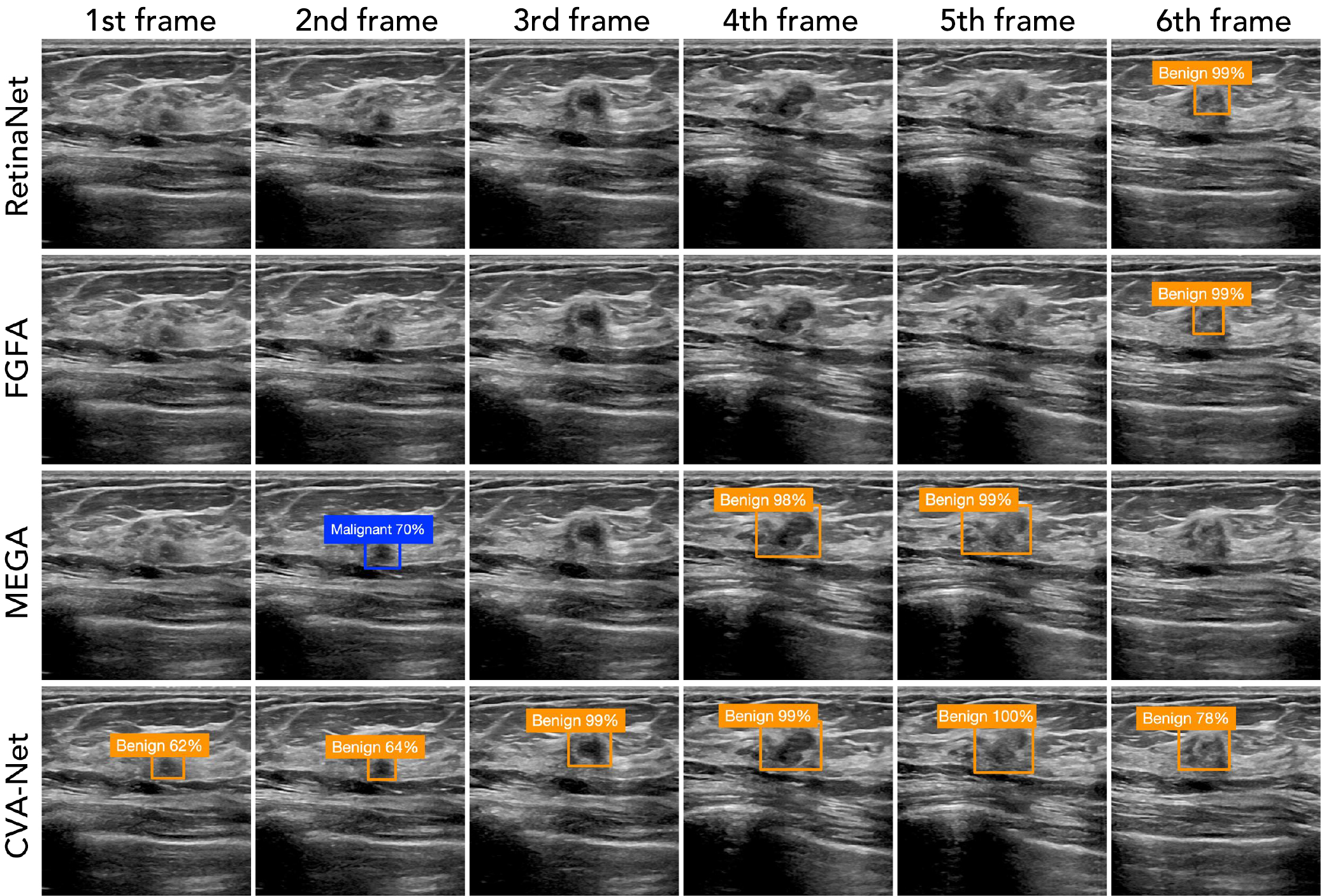}
\vspace{-3mm}
\caption{
  Visual comparisons of video breast lesion detection results produced by our network and state-of-the-art methods (i.e., RetinaNet, FGFA, and MEGA) for multiple frames of an ultrasound video with benign breast lesions.
}
\vspace{-3mm}
\label{fig:result}
\end{figure*}

\vspace{.5mm}
\noindent
\textbf{Visual Comparisons.} \
Fig.~\ref{fig:result} visually compares breast lesion detection results produced by our network and four compared methods for multiple frames of an ultrasound video with benign breast lesions.
As shown in the first two rows of Figure~\ref{fig:result}, RetinaNet and FGFA fail to detect the breast lesions of the first five frames.
MEGA obtains a better breast lesion detection result among three compared methods, but also fails to detect breast lesions at the 1st, 3rd, and 6th video frames.
Moreover, although MEGA can detect the breast lesion of the 2nd frame, the classification result is incorrect (see the blue rectangle of Fig.~\ref{fig:result}).
However, our method (see CVA-Net at last row of Fig.~\ref{fig:result}) can correctly detect the breast lesions of all video frames and has a correct breast lesion classification result for each video frame.

\begin{table}[!t]
\caption{
  Quantitative results of ablation study experiments.
  ``Ours-w/o-cla'' is to remove the lesion classification branch from our network.
  ``Ours-w/o-aug'' denotes that we do not utilize any data augmentation operation on shuffled video frames, while
  ``Ours-VF'', ``Ours-RR'', and ``Ours-CC'' denote that we utilize the vertical flip, random rotation, and center crop operation on the shuffled video frames.
}
\label{tab:ablation}
\centering
\begin{tabular}{l|ccc}
\bottomrule
 & AP & AP$_{50}$ & AP$_{75}$ \\
\hline
  Basic  & 27.8 & 52.8 & 26.4 \\
  Basic+Inter-video & 33.5 & 55.1 & 38.5 \\
  Basic+Intra-video & 27.8 & 60.9 & 36.3 \\
\hline
  Our method & \textbf{36.1} & \textbf{65.1} & \textbf{38.5} \\
\hline
  Ours-w/o-cla & 34.4 & 61.4 & 35.5 \\
\hline
  Ours-w/o-aug &35.6 & 62.1 & 37.8 \\
  Ours-VF & 35.5 & 59.8 & 37.1 \\
  Ours-RR & 35.4 & 62.1 & 38.4 \\
  Ours-CC & 35.8 & 62.4 & 37.9 \\
\toprule
\end{tabular}
\end{table}

\subsection{Ablation Study}

\noindent
\textbf{Effectiveness of clip-level features.} \
We first construct three baselines to verify the clip-level features learned by our inter-video fusion and intra-video fusion.
The first baseline network (``basic'') is to remove all inter-frame fusion modules and the intra-frame fusion modules from our network. 
The second baseline network (``basic+inter-video'') is to add inter-video fusion module into ``basic'', while the third one (``basic+intra-video'') is to add the intra-video fusion module into ``basic''.
As shown in Table~\ref{tab:ablation}, ``basic+inter-video'' has a larger AP, AP$_{50}$, and AP$_{75}$ scores than ``basic'', showing that the leveraging our inter-video fusion module to fuse local information of the input video and global information of the shuffled video can enhance the breast lesion detection performance in ultrasound video. 
Moreover, our network achieves superior results of AP, AP$_{50}$, and AP$_{75}$ over ``basic+inter-video'' and ``basic+intra-video''.
It improves the AP score from 27.8 to 36.1, the AP$_{50}$ score from 52.8 to 65.1,  and the AP$_{75}$ score from 26.5 to 38.6.
It indicates that integrating the inter-video fusion module and the intra-video fusion module together in our method can further enhance the breast lesion detection accuracy in ultrasound video.

\noindent
\textbf{Effectiveness of video-level features.} \ 
We also compare our method with and without the lesion classification branch and show their results in  Table~\ref{tab:ablation}.
It shows that the video breast lesion detection accuracy of our method are reduced when we remove the lesion classification branch from our network.

\noindent
\textbf{Effectiveness of data augmentation on shuffled video frames.} \
Note that we have utilized a data augmentation technique (i.e., random pepper) on shuffled video frames before passing them into the feature extraction backbone (see Encoder of Fig.~\ref{fig:pipeline}).
Hence, we further show the video breast lesion detection results of our network without and with different data augmentation techniques, including vertical flip, random rotation, and center crop.
Apparently, our method with a random pepper has the best detection performance, and thus our network empirically utilizes this data augmentation in our experiments. 

\section{Conclusion}
\label{sec:conclusion}

This paper first collects and annotates the first ultrasound video dataset for breast lesion detection with $188$ videos.
Moreover, we present a clip-level and video-level feature aggregation network for boosting breast lesion detection in ultrasound videos.
The main idea of our network is to combine clip-level features and video-level features for detecting breast lesions in videos, and the clip-level features are learned by devising inter-video fusion modules and intra-video fusion on the input ordered video and a shuffled video. 
Experimental results on our annotated dataset show that our network has obtained superior breast lesion detection performance over state-of-the-art methods.
Our further work includes the collection of more video data, exploration of a more systematic approach or complicated video shuffling operations.

\vspace{4mm}
\noindent
\textbf{Acknowledgment.}  This work was supported by the National Natural Science Foundation of China (No. 61902275, No. 12026604), AME Programmatic Fund (A20H4b0141), and Hong Kong Research Grants Council under General Research Fund (No. 15205919).

\bibliographystyle{splncs04}
\bibliography{paper1016.bib}

\end{document}